\documentclass[twocolumn,twoside,slac_two]{revtex4}
\usepackage[dvips]{graphicx}
\usepackage{amsmath}
\usepackage{fancyhdr}
\begin{document}

\title{Reach of Future Accelerator and Reactor Neutrino Efforts}

\author{Y. Obayashi for the T2K Collaboration}
\affiliation{Kamioka Observatory, Institute for Cosmic Ray Research, The University of Tokyo, GIfu, Japan}

\begin{abstract}
Numbers of accelerator and reactor neutrino oscillation experiments aiming to search
finite value of $\theta_{13}$ are starting within a few years.
T2K experiment is a next generation long baseline neutrino experiment starts on 2009, in Japan.
A main goal of the T2K experiment is to discover a finite $\theta_{13}$ by observing $\nu_e$ appearance.
Determining $\theta_{13}$ leads a next search of CP violation in the lepton sector.
Status of T2K experiment and prospects for CP measurement are reported in the report.
\end{abstract}

\maketitle

\thispagestyle{fancy}


\section{Introduction}
Neutrino mass and mixing matrix of the lepton sector can be explored only by neutrino
oscillations. Experimental results of the neutrino oscillations are indicating that
neutrino have finite mass and mixing matrix of the lepton sector is completely
different from of quark sector. 

Atmospheric and long baseline accelerator neutrino
experiments already measured $\sin^22\theta_{23}$ and $|\Delta m_{23}^2|$\cite{Fukuda94}.
Solar and long baseline reactor neutrino experiments have measured $\sin^2\theta_{12}$
\cite{Aharmim05}. However, as for the third mixing angle $\theta_{13}$,
only an upper limit has been determined\cite{Apollonio03}. Finite value of $\theta_{13}$
should be measured as soon as possible.

Numbers of next generation neutrino experiments using high intensity accelerator
or reactor are starting within a few years. These experiments are aiming to discover
the last unknown mixing angle $\theta_{13}$ by appearance ($\nu_\mu\rightarrow\nu_e$)
or disappearance ($\nu_e\rightarrow\nu_\mu,\nu_\tau$) oscillation channel. 

If finite $\theta_{13}$ is detected, size of the mixing could determine direction
of future experiments to search leptonic CP violation term. 

\section{T2K experiment}
T2K experiment is a 295km long baseline neutrino oscillation experiment
based on accelerator in Japan.
Muon neutrinos are produced by protons from 50 GeV synchrotron in J-PARC
striking on a 90cm long graphite target.
A large water Cerenkov detector Super-Kamiokande is used as a far detector.
The main physics goals of the experiment are precise measurement of
oscillation parameters, $\Delta m^2_{23}$ and $\sin^2 (2\theta_{23})$,
in $\nu_\mu$ disappearance channel and discovery of finite $\theta_{13}$
by observing $\nu_e$ appearance.

\subsection{Intense Narrow-band Neutrino Beam}
Neutrino beam is produced by $\pi^+\rightarrow\mu^+\nu_\mu$ decay from the
proton interaction in graphite target. Neutrino beam energy can be selected
by putting detector intentionally off-axis of parent $\pi^+$ direction.
We set the off-axis angle to $2.5^\circ$ in order to tune beam energy to be
0.6$\sim$0.7 GeV which corresponds that we can measure oscillation maximum
at 295km baseline when $\Delta m^2_{23}$ = (2.2 $\sim$ 2.6 ) $\times 10^{-3} eV^2$.

This off-axis method gives intense narrow-band neutrino beam which about
1,600 charged current interactions per year are expected in Super-Kamiokande.
It is two times larger intensity at the energy $\sim$0.7 GeV than the case of
``on-axis'' wide band beam. The off-axis beam has smaller high energy tail, and
it reduces background contamination for the energy reconstruction as described
below.

\begin{figure}[h]
\centering
\includegraphics[width=80mm]{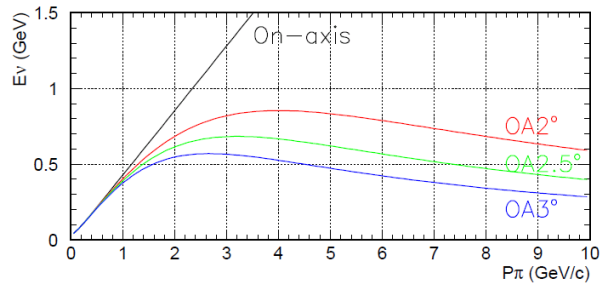}
\caption{Off-axis angle dependence of mean neutrino energy.} \label{offaxis}
\end{figure}

\begin{figure}[h]
\centering
\includegraphics[width=70mm]{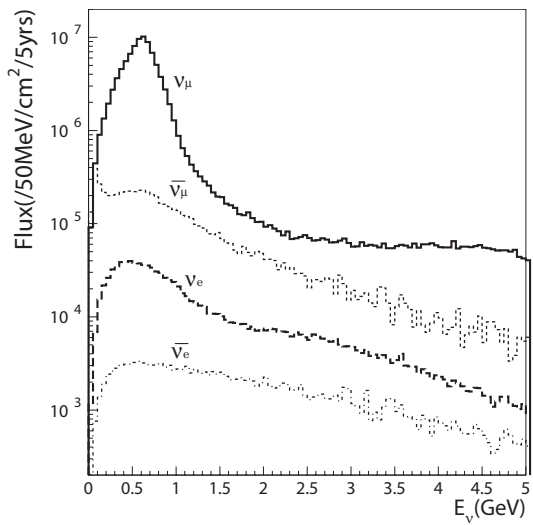}
\caption{Expected neutrino flux at 295km} \label{nuflux}
\end{figure}
\subsection{Near Detectors}
Properties of neutrino beam ``before'' oscillation and its interaction are measured by near detectors.
Near detectors consists of muon monitor placed at 110m downstream of proton target, and,
on-axis and off-axis neutrino detectors placed at 280m downstream of proton target.

The muon monitor measures direction and intensity of neutrino beam at each spill by monitoring
pion decay muons at just downstream of decay pipe. 

The on-axis detector monitors beam intensity, direction and profile of the neutrino beam.
The detector consists of 16 modules of iron and scintillator tracker blocks.
About 10k neutrino interactions per day is expected and measures beam direction within 1mrad accuracy.

The off-axis detectors are placed at $2.0 - 2.5^\circ$ off-axis region from the beam center
to measure neutrino beam goes to Super-Kamiokande. 
Detectors are surrounded by a large dipole magnet (re-use of UA1 magnet) which produces 0.2 T
uniform magnetic field. Upstream half consists of scintillator / Pb layers and 40\% of $H_2O$ target which dedicated to measure NC $\pi^0$ components. Downstream half consists of
two fine grained detectors that one is carbon target, another is $H_2O$ target, three TPC,
tracker calorimeter and side muon range detector. Iron of dipole magnet is also used as a
part of muon range detector. The detector measures off-axis beam flux, energy spectrum,
CC 1$\pi^+$, NC 1$\pi^0$ backgrounds, and used for neutrino cross-section studies. 

\begin{figure}[h]
\centering
\includegraphics[width=80mm]{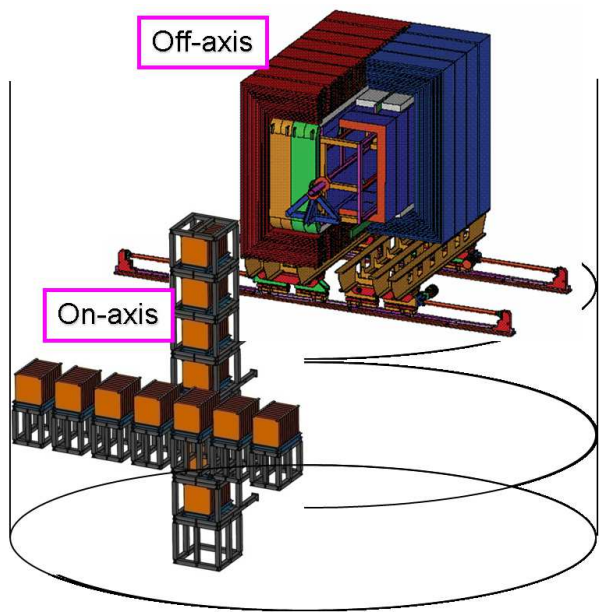}
\caption{Near detectors placed at 280m downstream of proton target} \label{nd280}
\end{figure}

\subsection{Far Detector: Super-Kamiokande}  

The far detector Super-Kamiokande is a cylindrically shaped water Cerenkov detector.
It consists 50 kton in total (22.5 kton fiducial volume) of pure water surrounded by 11127 20-inch photo
multiplier tubes. We are able to identify particle property using the feature
of Cerenkov ring image; shower-type ring for electron while sharp-edged ring
for muon. The muon-electron particle separation is better than 99\%.
The uncertainty of the energy scale is $\sim$2\%.
Detailed description of Super-Kamiokande detector can be found in \cite{Fukuda03}.

T2K neutrino events are selected from other background events by using timing
information. To enable the selection, both beam time stamp at J-PARC and event
time stamp at Super-Kamiokande are stamped by identical GPS systems. 
Beam timing information will be transfered to Super-Kamiokande immediately so that
real-time GPS timing reduction is enable.

\subsection{Physics Prospect}
\subsubsection{$\theta_{13}$: $\nu_e$ Appearance Measurement}

We measure $\nu_e$ appearance signal in $\nu_\mu$ beam by detecting Cerenkov ring of
single electron from $\nu_e$ CCQE interaction in Super-Kamiokande. 

Major backgrounds in the search are
intrinsic $\nu_e$ in the beam and mis-identification to single electron of single $\pi^0$
production in NC interactions. 
The $\nu_e$ component in the beam is 0.4\% at the peak of beam energy.
The NC $\pi^0$ is mis-identified when one of two photons from $\pi^0$ decay is mis-reconstructed
or two Cerenkov rings are overrapped.

In addition to the selection criteria for the single electron used in the Super-Kamiokande
atmospheric neutrino analysis, we apply specific $e/\pi^0$ separation cuts in order to 
further reduce NC $\pi^0$ background events. Then events which reconstructed neutrino
energy between 0.35 GeV and 0.85 GeV are selected.
Expected signals and backgrounds are shown in \tablename~\ref{table_nueapp}. 

\begin{figure}[h]
\centering
\includegraphics[width=80mm]{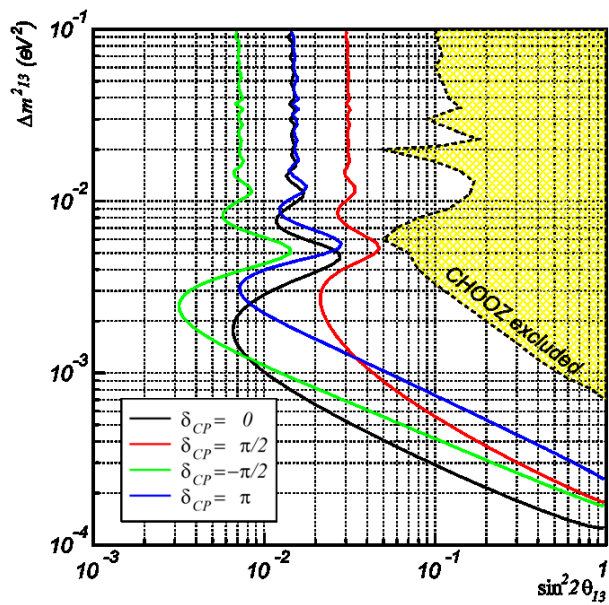}
\caption{Sensitivity of $\theta_{13}$ search as a function of $\Delta m^2_{13}$ in four cases of
$\delta_{CP}$ after T2K 5 years of exposure.} \label{nueapp}
\end{figure}

\figurename~\ref{nueapp} shows the expected sensitivity of $\theta_{13}$ as a function of
$\Delta m^2_{13}$. Even though dominant term is $\theta_{13}$ for
$\nu_\mu\rightarrow\nu_e$ appearance probability at 295km and neutrino energy of $\sim$
0.6 GeV, it is also depends on the other unknown parameter, $\delta_{CP}$.
In any case of $\delta_{CP}$, expected sensitivity of $\sin^22\theta_{13}$ is 10 times
or more smaller than current limit obtained by CHOOZ experiment around $\Delta m^2_{13} ~
2.5 \times 10^{-3} eV^2$.

\begin{table}[h]
\begin{center}
\caption{Expected signal and background events in $\nu_e$ appearance measurement after T2K 5 year exposure, assuming $\delta_{CP}=0$.}
\begin{tabular}{|c|c|c|c|c|}
\hline 
$\sin^2 2\theta_{13}$ & Signal & \multicolumn{3}{c|}{Background in Super-K}\\
\cline{3-5}
& &$\nu_\mu$ & Beam $\nu_e$ & Total \\
\hline
0.1 & 103 & 10 & 13 & 23 \\
\hline
0.01 & 10 & 10 & 13 & 23 \\
\hline
\end{tabular}
\label{table_nueapp}
\end{center}
\end{table}

\subsubsection{$\Delta m^2_{23}$, $\theta_{23}$: $\nu_\mu$ Disappearance Measurement}

We measure $\nu_\mu$ disappearance as both suppression in the total number of $\nu_\mu$
events observed at Super-Kamiokande and a distortion of the energy spectrum compared to 
the expected spectrum of no oscillation, which extraporated from the measurement at
production point.
We expect that our sensitivities for $\nu_\mu$ disappearance are $\delta (\Delta m^2_{23}) 
< 1 \times 10^{-4} eV^2$ and $\delta ( \sin^22\theta_{23} ) \sim 1\%$ with 5 year
of exposure.

\subsection{Construction Status}

Construction of accelerator complex in J-PARC is completed in 2008.
Commissioning of Linac and 3 GeV RCS ash succeeded in January and November 2007.
Then, commissioning of 50 GeV PS has succeeded in May 2008. 

Neutrino beamline has been constructed since 2004. Civil construction has been finished
and installation of super- and normal conducting magnets has been started. Manufacture and 
long term test of target and horn magnets are in progress and are on schedule.
Commissioning of the neutrino beamline will be in April 2009.

Neutrino detectors at J-PARC are under construction and will be installed in 2009.
Far detector Super-Kamiokande has been in operation since 1996. In 2008, DAQ electronics
and on-line computers will be upgraded and ready before neutrino beam starts.

\section{Future Accelerator and Reactor Experiments}
After the discovery of the neutrino oscillation and subsequential experiments, 
we have learned that a) neutrinos have small mass, b) mixing angles $\theta_{23}$
and $\theta_{12}$ are large, c) mixing angle $\theta_{13}$ is small (or zero).
But we still have many questions like: 
1) What is the value of third mixing angle $\theta_{13}$? 
2) $\theta_{23}$ is exactly 45$^\circ$ or not?
3) What is the ordering of the neutrino mass?
4) Is CP violated also in the lepton sector?
To find the answers of these questions, many accelerator and reactor neutrino oscillation
experiments are starting in these years in addition to T2K experiment.

\subsection{Accelerator Experiments} \label{Ex-acc}
Accelerator experiments measure $\theta_{23}$ and $\Delta m^2_{23}$ ($\sim\Delta m^2_{13}$)
with very high precision in $\nu_\mu\rightarrow\nu_x$ disappearance channel and searches
for $\theta_{13}$ in $\nu_\mu\rightarrow\nu_e$ appearance channel. 
MINOS experiment\cite{Gallagher08} with 735km baseline from Fermilab to Soudan started in May 2005 and disappearance measurement results are already published\cite{Adamson08}.
T2K experiment with 295km baseline from J-PARC and Kamioka starts physics run in 2009. 
By these long-baseline experiments, the mixing $\sin^22\theta_{23}$ and mass difference
will be pinned down and the third mixing $\sin^22\theta_{13}$ will be searched for
down to $\sim0.01$.
Another long-baseline neutrino oscillation experiment OPERA\cite{Rosa08} with 732km baseline from
CERN to CNGS started physics run in 2007. OPERA is designed to searches for 
$\nu_\mu\rightarrow\nu_\tau$ appearance channel by direct detection of $\nu_\tau$
interaction and expected to detect numbers of $\nu_\tau$
interactions within a few years of exposure. 

\subsection{Reactor Experiments} \label{Ex-reac}
Reactor experiments searches for $\theta_{13}$ in $\bar{\nu_e}\rightarrow\bar{\nu_{x}}$
disappearance channel. Double-CHOOZ experiment\cite{Lasserre08} in France will start physics run in 
middle 2009, Daya bay experiment\cite{White08} in China will start with all detector by end of 2010, and
RENO experiment\cite{White08} in Korea will start data taking in early 2010. All the experiments use
gadolinium doped liquid scintillator to detect $\bar\nu_e$ events and use same technique
that put near detector at $\sim$300m and far detector at $1\sim2$km from nuclear power
reactors.
Though expected deficit of $\bar{\nu_e}$ of reactor dis-appearance experiment is
quite small, this channel can measure $\theta_{13}$ free from $\delta_{CP}$
on contrary to the accelerator based appearance experiments. These reactor experiments
are expected to search $\sin^22\theta_{13}$ down to 0.01$\sim$0.03 after a few years
of running. 

\subsection{CP Measurement}

\begin{figure*}[hb]
\begin{equation*}
\begin{array}{rcl}
P(\overset{(-)}{\nu_\mu}\rightarrow\overset{(-)}{\nu_e})&
= & 4C^2_{13}S^2_{13}S^2_{23}\cdot\left( 1+\frac{2a}{\Delta m^2_{13}}(1-2S^2_{13})\right)
 \cdot\sin^2\Delta_{31}   \\
&+& 8C^2_{13}S_{12}S_{13}S_{23}(C_{12}C_{23}\cos\delta -S_{12}S_{13}S_{23})\cdot
 \cos\Delta_{32}\cdot\sin\Delta_{31}\cdot\sin\Delta_{21}    \\
&\pm& 8C^2_{13}C_{12}C_{23}S_{12}S_{13}S_{23}\sin\delta\cdot\sin\Delta_{32}
 \cdot\sin\Delta_{31}\cdot\sin\Delta_{21} \\
&+&4\S^2_{12}C^2_{13}(C^2_{12}C^2_{23}+S^2_{12}S^2_{23}S^2_{13}-
 2C_{12}C_{23}S_{12}S_{23}S_{13}\cos\delta )\cdot\sin^2\Delta_{21}\\
&-&8C^2_{13}S^2_{13}S^2_{23}\cdot\frac{aL}{4E_{\nu}}(1-2S^2_{13})\cdot
 \cos\Delta_{32}\cdot\sin\Delta_{31}

\end{array}
\end{equation*}
\caption{Oscillation probability of $\nu_\mu\rightarrow\nu_e$
($\bar{\nu_\mu}\rightarrow\bar{\nu_e}$) where $C_{ij}=\cos\theta_{ij}$,
$S_{ij}=\sin\theta_{ij}$, $\Delta_{ij}=\Delta m^2_{ij}\cdot L/4E_\nu$,
$L$ is neutrino flight length in km, $E_\nu$ is neutrino energy in GeV,
$a=2\sqrt{2}G_FN_eE_\nu$ and $\delta$ is CP phase. }
\label{oscprob}
\end{figure*}

\begin{figure*}
\centering
\includegraphics[width=80mm]{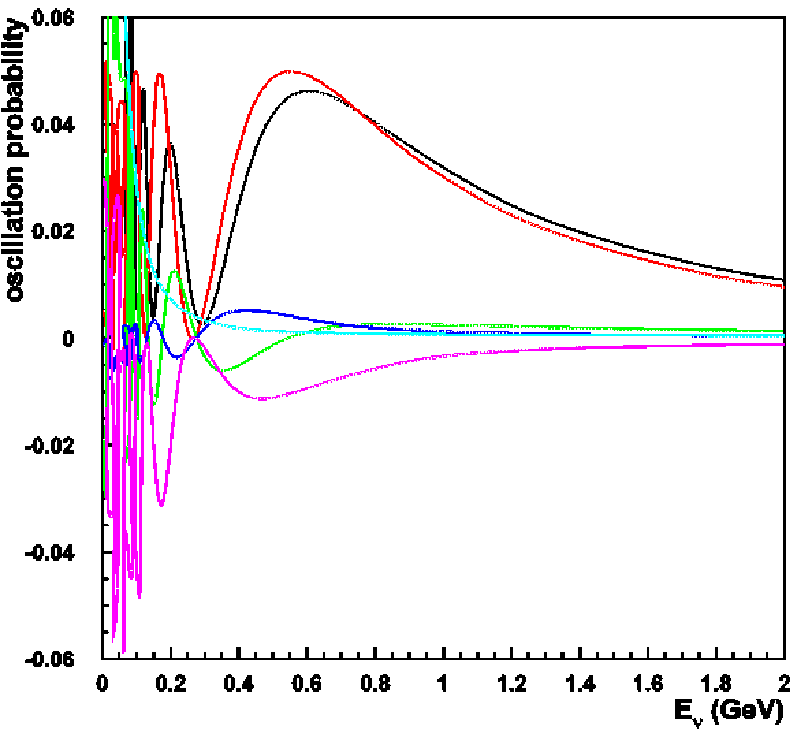}
\includegraphics[width=80mm]{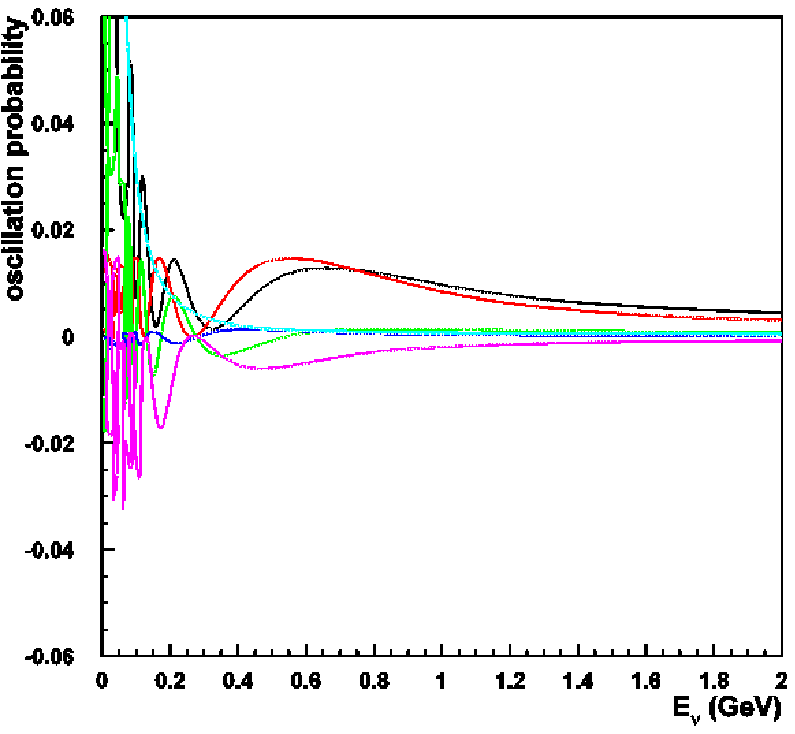}

\caption{Oscillation probability in the function of neutrino energy $E_\nu$
at neutrino flight length $L$=295km, CP phase $\delta=\pi /4$, $\sin^22\theta_{23}=1$
and $\sin^22\theta_{13}$=(left:)0.1, (right:)0.03.
Black lines show total oscillation probability, 
red, cyan, magenta, green and blue lines shows each component of first,
second, third, fourth and fifth term in \figurename~\ref{oscprob} respectively. }\label{oscfig}

\end{figure*}

As a future expectation of early 2010's,
neutrino oscillation experiments mentioned in previous section will get
knowledge about $\theta_{13}$.
If we could get finite $\theta_{13}$ value, our target will
change to next question 3) and 4). On the other hand, if we could get
just small upper limit of $\sin^22\theta_{13}$, we still need to explore
further small value. 
In any case, the measurement will become far precise
and need huge statistics. Neutrino oscillation probability including
CP phase $\delta$ and matter effects in the earth is shown in 
\figurename~\ref{oscprob}. Two examples of oscillation probability in
the case of $\sin^22\theta_{13}=$0.1(just below CHOOZ boundary) and 0.03(around
sensitivity limit of experiments shown in previous subsections) are shown in
\figurename~\ref{oscfig}.

To realize such a precise measurement with huge statistics, in addition to
the improvement of
accelerator power, several techniques of huge neutrino detectors are under study.
Hyper-Kamiokande\cite{Okumura07} is a gigantic water Cerenkov detector under study to be built
in Kamioka. Here, we assume J-PARC accelerator power is upgraded to 1.7MW and expose neutrino
beam to Hyper-Kamiokande at the same length of baseline with T2K.
Expected $\nu_\mu\rightarrow\nu_e$ signal and backgrounds are shown in Table~\ref{cpsearch}.
If $\delta_{CP}$ is non-zero value, $\nu_\mu\rightarrow\nu_e$ oscillation probability is
differ from $\bar{\nu_\mu}\rightarrow\bar{\nu_e}$ case. We can search $\delta_{CP}$ value by
comparing the oscillation probabilities of neutrino and anti-neutrino case.
Since the interaction cross-section with proton and neutron is smaller for anti-neutrino,
we need longer exposure time for anti-neutrino run.

As another possible detector to explore $\delta_{CP}$, large liquid argon TPC detector is
under study\cite{Maruyama08}. By measuring energy spectrum of lower energy
neutrino than water Cerenkov threshold, we can measure $\delta_{CP}$ from a spectrum
distortion of second oscillation maximum without anti-neutrino run.
To measure it, longer baseline than current T2K baseline(295km) should be optimum.
The study \cite{Maruyama08} suggests Okinoshima island is a possible site.

Before moving into CP measurement, we need to have knowledge about
$\theta_{13}$. Getting results about $\theta_{13}$
by the experiments mentioned in subsections \ref{Ex-acc} and \ref{Ex-reac}
as early as possible is one of most important requirement for neutrino
oscillation physics in next $\sim$5 years.

\begin{table}
\caption{expected $\nu_\mu\rightarrow\nu_e$ signal and backgrounds for future $\delta_{CP}$
search assuming 2.2year $\nu_\mu$ and 7.8year $\bar{\nu_\mu}$ exposure by upgraded 1.7MW
accelerator at J-Parc to 0.54 Mega-ton fiducial volume water Cerenkov detector placed
at Kamioka.\label{cpsearch}} 
\begin{tabular}{|c|c|c|c|c|c|c|}
\hline
&\multicolumn{2}{c|}{$\nu_\mu\rightarrow\nu_e$ signal}&\multicolumn{4}{c|}{Background}\\
\hline
&$\delta_{CP}=0$&$\delta_{CP}=\pi /2$&$\nu_\mu$&$\bar{\nu_\mu}$&$\nu_e$&$\bar{\nu_e}$\\
\hline
$\nu_\mu$ 2.2yr & 1049 & 579 & 354 & 26 & 379 & 10 \\
\hline
$\bar{\nu_\mu}$ 7.8yr & 1050 & 1493 & 443 & 610 & 241 & 415 \\
\hline
\end{tabular}
\end{table}

\begin{figure*}
\centering
\includegraphics[width=90mm]{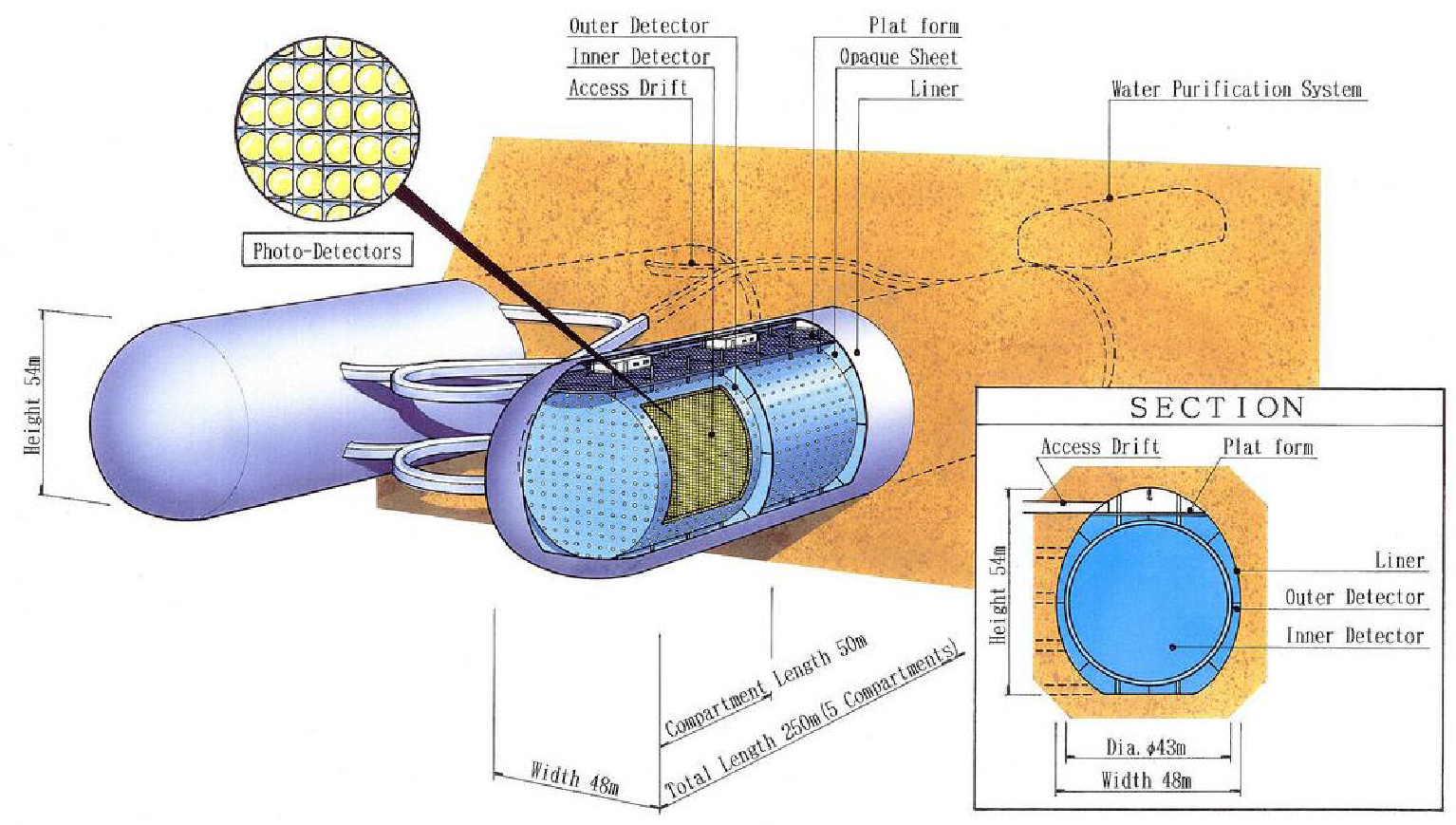}

\caption{Schematic view of gigantic water Cerenkov neutrino detector Hyper-Kamiokande.
Planning total and fiducial volumes are 1 Mega-ton and 0.54 Mega-ton, respectively.
See \cite{Okumura07} in detail.}\label{hyperk}

\end{figure*}

\section{Summary}

Next generation accelerator and reactor neutrino experiments aiming to measure
finite value of $\theta_{13}$ are starting in a few years. Long baseline accelerator
experiment T2K will start on Apr. 2009. If non-zero $\theta_{13}$ is observed,
oscillation experiments should proceed to the next phase. The main goal of the
next phase is the observation of CP violation in the neutrino sector. 
For the observation, we need precise measurement with huge statistics.
The measurement of $\theta_{13}$ as soon as possible is necessary to determine
the direction of the next phase experiments.

\begin{acknowledgments}
The author thanks the organizers of the Flavor Physics \& CP 
Violation (FPCP) 2008 conference for the invitation to present
this talk and hospitality of the conference.
This work was partly supported by the Grant-in-Aid for Scientific Research, Japan Society for the Promotion of Science.

\end{acknowledgments}

\bigskip 

\end{document}